\documentclass[10pt,letterpaper]{article}
\usepackage[top=0.85in,left=1.in,footskip=0.75in,marginparwidth=2in]{geometry}

\usepackage[utf8]{inputenc}

\usepackage{cite}

\usepackage{nameref,hyperref}

\usepackage[right]{lineno}

\usepackage{microtype}
\DisableLigatures[f]{encoding = *, family = * }

\raggedright
\usepackage{changepage}

\usepackage[aboveskip=1pt,labelfont=bf,labelsep=period,singlelinecheck=off]{caption}

\makeatletter
\renewcommand{\@biblabel}[1]{\quad#1.}
\makeatother

\usepackage{lastpage,fancyhdr,graphicx}
\usepackage{epstopdf}
\fancyheadoffset[L]{2.25in}
\fancyfootoffset[L]{2.25in}

\usepackage{color}

\definecolor{Gray}{gray}{.25}

\usepackage{graphicx}

\usepackage{sidecap}

\usepackage{tabularx}
\usepackage{mathrsfs}
\usepackage{lineno}
\usepackage{epsfig}
\usepackage{amsmath}
\usepackage{amssymb}
\usepackage{float}
\usepackage{enumerate}
\usepackage{lineno}
\usepackage{booktabs}
\usepackage{arydshln}

\usepackage{wrapfig}
\usepackage[pscoord]{eso-pic}
\usepackage[fulladjust]{marginnote}
\reversemarginpar

\begin{document}
\vspace*{0.35in}

\begin{flushleft}
{\Large
\textbf\newline{Electrical Characteristics of the GEC Reference Cell with Impedance Matching: A Two-Dimensional PIC/MCC Modeling Study}
}
\newline
\\
Zili Chen \textsuperscript{1},
Hongyu Wang \textsuperscript{2,*},
Shimin Yu \textsuperscript{1},
Yu Wang \textsuperscript{3},
Zhipeng Chen \textsuperscript{1},
Wei Jiang \textsuperscript{3},
Julian Schulze \textsuperscript{4,*}
Ya Zhang \textsuperscript{5,*}
\\
\bigskip
\bf{1} State Key Laboratory of Advanced Electromagnetic Technology, International Joint Research Laboratory of Magnetic Confinement Fusion and Plasma Physics, School of Electrical and Electronic Engineering, Huazhong University of Science and Technology, Wuhan, 430074, China
\\
\bf{2} School of Physics Science and Technology, Anshan Normal University, Anshan 114007,China
\\
\bf{3} School of Physics, Huazhong University of Science and Technology, Wuhan 430074,China
\\
\bf{4} Chair of Applied Electrodynamics and Plasma Technology, Faculty of Electrical Engineering and Information Sciences, Ruhr University Bochum, 44801 Bochum, Germany
\\
\bf{5} Department of Physics, Wuhan University of Technology, Wuhan 430070, China
\\
\bigskip
* wanghy\_asnc@hotmail.com, yazhang@whut.edu.cn

\end{flushleft}

\section*{Abstract}
In this paper, the electrical characteristics of the Gaseous Electronics Conference (GEC) reference cell with impedance matching are investigated through a two-dimensional electrostatic implicit Particle-in-Cell/Monte Carlo Collision (PIC/MCC) model in an axisymmetric coordinate system. The coupling between the complex reactor geometry and the external circuit is included via an equivalent capacitance calculated from the electric energy density. The results of this model are compared with experimental measurements and other model calculations and show good agreement. This simulation obtains the plasma kinetics of the capacitively coupled discharge process at low pressure and detailed external circuit responses, including power transmission, reflection, and higher-order harmonics in the circuit, which provides important insights for impedance-matching design in semiconductor plasma processing.


\section{Introduction}

Low-temperature plasmas (LTPs) such as capacitively coupled plasmas (CCPs) and inductively coupled plasmas (ICPs) are widely used in microelectronics manufacturing for material processing, particularly in etching and deposition. Predictive modeling of these plasmas is always desired for industrial and laboratory reactor design and to gain insight into the fundamental physics of discharge processes \cite{lieberman_principles_2005, chabert_physics_2011}.

Various simulation techniques have been utilized in recent years to predict the behavior of LTPs, including fluid, Particle-in-Cell/Monte Carlo Collision (PIC/MCC) and hybrid models. One-dimensional models \cite{vahedi_capacitive_1993, verboncoeur_simultaneous_1993, kawamura_stochastic_2006, sahu_full_2020, yu_generalized_2023} have been commonly used due to their lower computational cost, but they offer limited information. On the contrary, two-dimensional models have gained more attention because they consider actual chamber geometries. Fluid models, which solve conservation equations (mass, momentum, energy and higher moments) under the assumption of a Maxwellian velocity distribution, are one of the principal methods \cite{sahu_full_2020, alvarez_laguna_asymptotic_2020, baldry_continuum_2021}. SOMAFOAM \cite{verma_somafoam_2021}, an OpenFoam-based finite volume framework, is a typical representative that allows large-scale parallel modeling of arbitrary discharge chambers. Fluid models are generally suitable for modeling discharge processes at high pressure. Kinetic models, including PIC/MCC methods and direct kinetic simulations \cite{hara_test_2018, liu_conservative_2021}, are necessary when the pressure decreases and the mean free path becomes longer. This method is generally only suitable for small-scale simulation, due to the high computational cost, as the Debye length, cyclotron radius, and plasma oscillation frequency must be resolved. PIC/MCC models based on graphics processing units (GPUs) \cite{kim_advanced_2018, hur_model_2019, kim_two-dimensional_2022} and multicore central processing units (CPUs) \cite{wang_parallelization_2009} have been developed to improve computing efficiency. The computational efficiency of the PIC/MCC models based on implicit push algorithms \cite{vahedi_capacitive_1993-1, kawamura_physical_2000, wang_implicit_2010, bai_implicit_2021, eremin_energy-_2022} has been improved by orders of magnitude due to the smaller space and time steps explicitly allowed \cite{wakayama_study_2003, takekida_particle_2005, charoy_2d_2019, villafana_2d_2021, kim_advanced_2018, hur_model_2019, kim_two-dimensional_2022}. To simulate complex geometric chambers, picFoam \cite{kuhn_picfoam_2021} based on the OpenFoam frame uses the finite volume method to achieve a fully kinetic electrostatic PIC/MCC model, which has good extendability and can simulate arbitrary geometries in 1 to 3 dimensions. Hybrid models \cite{kushner_hybrid_2009} are also an effective way of dealing with different physical phenomena on different time scales.

The discharge characteristics of the chamber are highly sensitive to the configuration of the external circuit due to the nonlinear interaction between the circuit and the plasma \cite{hargis_gaseous_1994, olthoff_gaseous_1995}. Small changes in circuit parameters, such as cable length, matching network parameters, and radio frequency source (RF) configuration, can significantly affect discharge states. Therefore, describing the external circuit and plasma simultaneously in LTP simulations becomes necessary. This approach enables the self-consistent description of the discharge characteristics of LTPs. In addition, studying external circuit considerations can facilitate the research of impedance matching, a crucial aspect of semiconductor plasma processing. By designing the input impedance of the load or the output impedance of the RF source, impedance matching can effectively improve power transport and minimize power reflection. Impedance matching can maximize power transmission and prevent cable damage due to excessive reflected power. It can also make the plasma absorb the same power when operating conditions change, ensuring the repeatability of the discharge process \cite{zhang_impedance_2012}.

Impedance matching poses a persistent challenge in semiconductor plasma processing. Various impedance matching networks (IMN) with different structures \cite{wang_experimental_2019, he_improvement_2023} and tuning algorithms \cite{zhang_impedance_2012} have been designed for automatic impedance matching. Since plasma impedance is difficult to measure, extremum seeking control methods have become the practical industrial standard. However, a practical problem is that the range of variable capacitors still needs to be determined \cite{zhang_impedance_2012}. In general, the tuning space needs to be given by some numerical models or based on trial-and-error methods. Additionally, numerical models can help investigate the impact of impedance matching on discharge characteristics and deepen their understanding \cite{qu_power_2020}.

The above studies have developed discharge models for various reactor geometries that explain the kinetics of LTPs. However, these models typically do not self-consistently include the external impedance matching. On the other hand, experimental voltage, current, and impedance measurements are always a tough task, as disturbance of the probes and thus complex compensation method will be required. Indeed, the electrical characteristics are still not clear for GEC reactors, altrough great efforts have been taken over decades \cite{sobolewskiElectricalCharacteristicsArgon1995,spiliopoulosPowerDissipationImpedance1996}.

In this work, we propose an implicit electrostatic PIC/MCC model that can describe capacitively coupled discharges generated in a gaseous electronics conference (GEC) reference cell in a two-dimensional cylindrical coordinate system that self-consistently includes the external impedance matching. Our model can be solved for arbitrary external circuits and multiple electrodes of any shape simultaneously. To verify our proposed model, we compared it with published kinetic calculations \cite{rauf_uniformity_2020} and experimental results \cite{hargis_gaseous_1994, overzet_microwave_1995}. Furthermore, we investigated the discharge process at different pressures, considering the effects of the impedance matching. The computational model is described in Sec.\ref{sec2}. The results of model validation and calculation are discussed in Sect.\ref{sec3}. The conclusion is drawn in Sec.\ref{sec4}.

\section{Computational model} \label{sec2}
The implicit electrostatic PIC/MCC model presented in this paper is a two-dimensional simulation based on our previous research, which comprises one-dimensional \cite{wang_energy_2014, wu_electrical_2021}, two-dimensional \cite{wang_parallelization_2009, wang_implicit_2010, jiang_implicit_2011}, and recently published external circuit models \cite{yu_generalized_2023, chen_numerical_2022}. The PIC/MCC algorithm employed in this model aligns with the ones used previously, but incorporates a complex reactor geometry module and an external circuit module to capture the generation of a capacitively coupled plasma realistically. The simulation is carried out at low pressure and does not take into account secondary electron emission, field emission, thermal emission, and recombination processes, which could, however, easily be added based on our prior work \cite{wu_electrical_2021, guo_numerical_2022, zhong_numerical_2022}.

\begin{figure*}[ht]
    \centering
    \includegraphics[width=\linewidth]{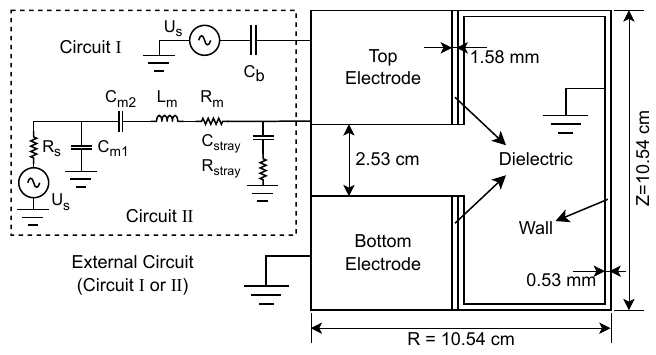}
    \caption{The schematic of the GEC reference cell with two external circuits.}
    \label{fig: 1}
\end{figure*}

The GEC reference cell is chosen as the simulated discharge chamber to facilitate comparison with the experiment and other numerical calculation results, as shown in Fig.\ref{fig: 1}. The GEC reference cell comprises a top electrode, a bottom electrode, a wall, and dielectrics. The top electrode is connected to a radio frequency (RF) drive circuit, while the bottom electrode and the wall are grounded. The dielectrics provide insulation between the metal electrode and the wall. The chamber configuration is mainly based on the literature \cite{rauf_uniformity_2020, hara_effects_2023}. Taking into account the simple regular grid layout commonly used in PIC/MCC simulation, the GEC reactor size is slightly modified in our simulations. The radius and height of the chamber are 10.54 cm, which are discretized into $200 \times 200$ grid cells. The electrode spacing is 48 grid cells, which is 2.53 cm. The wall thickness is one grid division, which is 0.53 mm. The relative dielectric constant of the dielectrics is 12. 

Here, two different external circuits are considered. External circuit I is a simple configuration consisting of only an RF source and a blocking capacitor. This circuit is used in many simulations \cite{rauf_uniformity_2020} and is used mainly to verify our two-dimensional PIC/MCC model by comparing its results with experimental and other simulation results. External circuit II is a typical configuration that considers impedance matching, which has been used in our previous study of one-dimensional models \cite{yu_best_2022, chen_numerical_2022}. The power generated by the RF source is applied to the discharge chamber of the GEC reference reactor through an L-type impedance matching network (composed of a parallel capacitor and an RLC series branch) to excite the plasma. A stray branch is also taken into account.

To simulate the reactor geometry of the GEC reference cell, we used the electrostatic field solution of the axisymmetric PIC/MCC model, taking into account both metal and dielectric materials. The Poisson equation describes the heterogeneous system under the implicit scheme, which is expressed as follows:
\begin{equation}
 \nabla\cdot[(1+\chi(z,r)) \epsilon(z,r) \nabla\phi]=-\tilde{\rho}(z,r)
\label{eq: poisson implicit}
\end{equation}
where, $\tilde{\rho}$ denotes the charge density of particles after moving only based on the current time step, while $\phi$ represents the space potential. $\chi$ is the relative permittivity of the materials or the numerical correction factor in the implicit scheme. $\epsilon$ is the dielectric constant. As the frequency is low, the metal is considered an equipotential body, meaning that the potential of the top electrode, the bottom electrode, and the wall is fixed as the boundary condition of the Poisson equation. When coupled to the external circuit, the original Verboncoeur's method \cite{verboncoeur_simultaneous_1993} requires iterations and is computationally extensive. When multiple electrodes are connected to different external circuits, it becomes necessary to utilize Vahedi's method \cite{vahedi_simultaneous_1997} to decompose the spatial potential into the solution $\phi_P$ of the Poisson equation (Eq.\ref{eq: poisson zero}) with zero boundary conditions and the solutions $\phi_{Li}$ of multiple Laplace equations (Eq.\ref{eq: Laplace}) with normalized boundary conditions.

\begin{equation}
\phi = \phi_p + \sum_i U_{Ei} \phi_{Li}
\label{eq: vahedi}
\end{equation}
\begin{equation}
\left\{
\begin{array}{l}
\begin{aligned}
\nabla^2 &\phi_P = - \frac{\rho}{\epsilon_0}\\
&\phi_b = 0\\
\end{aligned}
\end{array}
\right.
\label{eq: poisson zero}
\end{equation}

\begin{equation}
\left\{
\begin{array}{l}
\begin{aligned}
\nabla^2 &\phi_{Li} = 0\\
&\phi_b = 1\\
\end{aligned}
\end{array}
\right.
\label{eq: Laplace}
\end{equation}
where the field boundary is denoted by $\phi_b$, $U_{Ei}$ is the voltage on the electrode. Note that for explicit PIC schemes, the solution of the Laplace equation is invariant and only needs to be solved once during program initialization. For the implicit PIC scheme used here, due to the change in the numerical correction factor, the Laplace equation must be solved at every step for each electrode. However, we have found that $\chi$ is very low and does not change much in most cases, so it is sufficient to determine it once per RF period. Hara\cite{hara_effects_2023} has presented an improved version of the Vahedi method described above with great success. However, it still requires the sum of all the electric fields over the entire electrode. 

Here, we adopted a simpler and more effective method proposed by Wang et al.\cite{wanghongyu_2017_conference}. This method allows the PIC/MC method to include multiple electrodes that couple to different external circuits of any arbitrary complexity. In Wang's method, the potential $U_E$ on each metal electrode is obtained by the accumulated charge $Q_E$ on the electrode and the equivalent capacitance $C_{eq}$ of the metal electrode. 

\begin{equation}
U_E = \frac{Q_E}{C_{eq}}
\label{eq: cap}
\end{equation}

The equivalent capacitance of each metal electrode is determined via electric energy. The potential of a metal electrode is set at U$_E$ = 1 V, while the other electrodes remain grounded. At this point, the electrostatic solver is deployed to compute the vacuum electric field $\mathbf{E}$, following which the electric field energy density $u_e$ is obtained according to the following expression:
\begin{equation}
u_e = \frac{1}{2} \epsilon |\mathbf{E}|^2
\label{eq: ue}
\end{equation}

As such, the equivalent capacitance can be calculated in the following form:
\begin{equation}
\mathscr{E}_e = \frac{1}{2} C_{eq} U_E^2 = \int u_e dV
\label{eq: Ceq}
\end{equation}
where $\mathscr{E}_e$ represents the total energy of the electric field. The Poisson equation can be solved directly when only one electrode is connected to the external circuit (with the other electrodes remaining grounded). 

The equivalent capacitance determines the electrode potential, after which the plasma can be coupled with the generalized Verboncoeur method\cite{verboncoeur_simultaneous_1993, yu_generalized_2023}. At each time step, the plasma and external circuit models are solved simultaneously. The charge conservation equation is used to obtain the charge $Q_E$ on each electrode by:
\begin{equation}
\frac{d Q_E}{d t} = I_{ccp} + I_{conv},
\label{eq: charge conservation}
\end{equation}
where $ I_{ccp}$ is the current flowing from the external circuit to the attached electrode, and $I_{conv}$ is the convective current from plasma to electrode. Next, the potential at the electrode $U_E$ is determined by the charge $Q_E$ and the equivalent capacitance calculated by the electric energy density. The potential can be used as a boundary condition to solve the differential equations of the external circuit and the Poisson equation.

In circuit I, a more general branch of the RLC series replaces the blocking capacitor $C_b$. After Simulink \cite{the_mathworks_simscape_2022} verification, the circuit response is almost the same when the blocking capacitor is connected in series with a small resistance $R_b = 0.5 ~\Omega$ and inductance $L_b = 0.01 ~\mu H$, but better convergence is obtained. The differential equations for circuit I are given by:
\begin{equation}
\left\{
\begin{array}{l}
\begin{aligned}
\frac{d Q_{b}}{dt} &= I_{b}\\ 
\frac{d I_{b}}{dt} &= \frac{1}{L_b} \left[ U_s - U_E - \frac{Q_b}{C_b} - I_b R_b \right]\\
\end{aligned}
\end{array}
\right.
\label{eq: RLC}
\end{equation}
where $I_b$ and $Q_b$ are the current and charge of the loop where the capacitor $C_b$ is located, respectively. $U_s$ is the source voltage. 

In circuit II, the current and charge of the loop where the power supply is located are $I_1$ and $Q_1$, respectively. The current and charge of the loop where $C_{m2}$ is located are $I_{2}$ and $Q_{2}$, respectively. The current and charge of the loop where the plasma load is located are $I_{ccp}$ and $Q_{ccp}$, respectively. The voltage of the plasma load is $U_{ccp}$. The current flowing through the stray capacitance is $I_{stray}$. The circuit differential equations (similarly, a small inductance $L_{stray}$ is added to the stray branch to give differential equations without affecting the circuit response) for circuit II are as follows:
\begin{equation}
\left\{
\begin{array}{l}
\begin{aligned}
\frac{d Q_{1}}{dt} &= \frac{1}{R_s} \left[U_s - \frac{Q_1-Q_2}{C_{m1}} \right] \\
\frac{d Q_2}{dt} &= I_2\\
\frac{d I_{2}}{dt} &= \frac{1}{L_m} \left[ \frac{Q_1-Q_2}{C_{m1}} - U_{ccp} - \frac{Q_2}{C_{m2}} - I_2 R_m \right]\\
\frac{d Q_{ccp}}{dt} &= I_2 - I_{stray}\\
\frac{d I_{stray}}{dt} &= \frac{1}{L_{stray}} \left[U_{ccp} - \frac{Q_2-Q_{ccp}}{C_{stray}} - I_{stray} R_{stray} \right]\\
\end{aligned}
\end{array}
\right.
\label{eq: IMN}
\end{equation}

In the cylindrical coordinate system, because of the difference in cell volume in the radial direction, unequal-weight macroparticles must be used to reduce the number of macroparticles used. However, this approach presents a new challenge in the form of nonphysical heating caused by noise on the central axis. The presented work addresses this problem by implementing particle splitting, which follows our previous strategy \cite{wang_implicit_2010}. A better treatment strategy can refer to Kentaro's work \cite{hara_effects_2023}. To mitigate the significant increase in the number of macroparticles associated with particle splitting, we first simulate the steady state without splitting, and then introduce the splitting to reach the steady state again. The number of macroparticles ranges from 2 to 5 million before splitting. After the implementation of particle splitting, the number of particles stabilizes between 25 million and 50 million. In fact, this non-physical heating can be significantly improved by increasing the number of particles per cell\cite{turner_kinetic_2006}, decreasing the space and time steps or adopting the energy-conserving scheme \cite{wang_energy_2014}. 

Collision reactions are considered using Vahedi's null-collision method \cite{vahedi_monte_1995}, which has been widely used in our previous studies, including inert Ar gas \cite{wu_electrical_2021}, electronegative gases such as CF4 \cite{wu_breakdown_2022}, SF6 \cite{gao_computational_2021}, etc. This study focuses on impedance matching of the GEC reference cell rather than chemical reactions; hence, the use of the inert gas Ar with simple collision processes. The simulation considers only electrons, single ionized ions, and neutral atoms. The simulation is carried out at low pressure (<100 mTorr) so that the plasma ionization rate is very low and the change in the background gas density is not considered. Neutral atoms are sampled from the Maxwellian distribution at 300 K during collision processes. The simulation contains collision reactions between electrons and neutral atoms, including elastic, ionization, and excitation collisions, including elastic collisions, and charge exchange between single-ionized ions and neutral atoms.

Although most of the two-dimensional PIC/MCC model algorithms applied in this work are derived from our previous work, the code has been completely redesigned and modernized. Fortran 2018 standard is used to improve the readability, maintainability, and expansibility of the code. Additionally, we utilized HDF5 \cite{hdf5} to speed up data storage, and SUNDIALS \cite{hindmarsh2005sundials} and PETSc \cite{osti_1614847} are used to solve the external circuit equations and the Poisson equation, respectively. To handle the massive computation required by the two-dimensional PIC/MCC model, domain decomposition is used for parallelization, with either the Fortran 2018 Coarray or MPI 3.1 \cite{gropp_using_2014} standard.

\section{Results} \label{sec3}
In the simulation, the capacitively coupled discharge operated in the GEC reference cell is initialized with a uniform plasma density profile at low pressure. The initial electron number density is $1 \times 10^{14} ~\rm m^{-3}$, and the initial electron temperature is 30000 K. The implicit push scheme allows setting the time to $5 \times 10^{-10} ~\rm s$, and the initial number of macroparticles per grid cell is set to 50. The frequency of the RF source is fixed at 13.56 MHz. In circuit I, the blocking capacitor is set to 5 nF, consistent with the configuration of Rauf's model \cite{rauf_uniformity_2020}. In circuit II, the internal resistance $R_s$ of the power supply is set to 50 $\Omega$. The loss of the impedance matching network is represented by $R_m$ = 0.5 $\Omega$, and the loss of the stray branch by $R_{stray}$ = 0.5 $\Omega$. According to vendor information, the stray capacitance $C_{stray}$ is set at 10 pF, and the inductance $L_m$ is set at 0.85 $\mu \rm H$. The equivalent capacitance $C_{eq}$ of the vacuum GEC reference cell is 892.07 pF, obtained by the energy method. The equivalent capacitance replaces the GEC reference cell, and Yu's method \cite{yu_best_2022} calculates the matching parameters to be $C_{m1}$ = 2335.51 pF and $C_{m2}$ = 215.62 pF for perfect impedance matching under vacuum condition.

\begin{figure*}[ht]
    \centering
    \includegraphics[width=\linewidth]{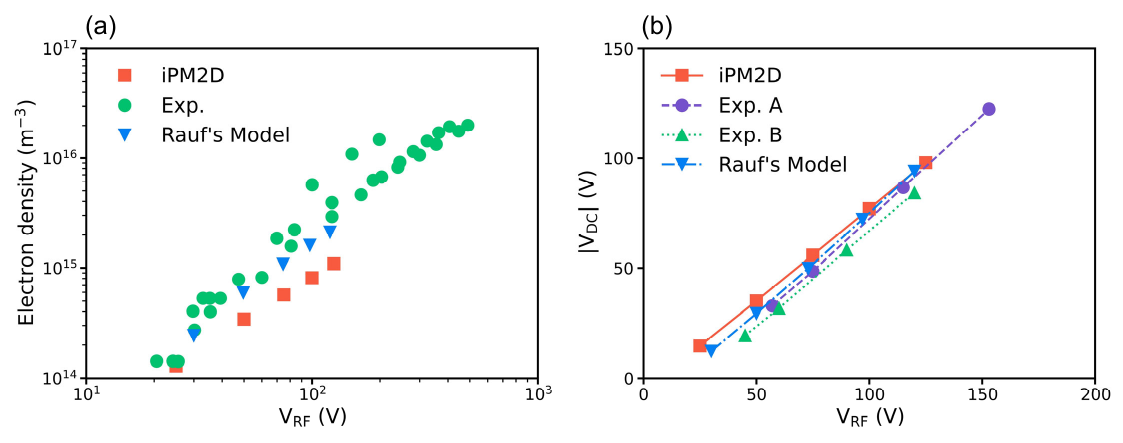}
    \caption{At 100 mTorr, the average electron number density (a) and DC self-bias voltage (b) corresponding to different RF voltages obtained by iPM2D, experimental measurement (electron density from microwave interferometry \cite{overzet_microwave_1995}, self-bias voltage obtained with resonance at 24 MHz, Exp. A, and 34 MHz, Exp. B \cite{hargis_gaseous_1994}) and Rauf's model \cite{rauf_uniformity_2020}.} 
    \label{fig: 2}
\end{figure*}

Circuit I is used to validate the proposed computational model, iPM2D. Due to the large blocking capacitance, a direct current (DC) self-bias voltage is gradually established. To ensure that the discharge reaches a steady state, the simulation is run for 9000 RF cycles, of which the last 2000 cycles use particle splitting. The average electron number density and DC self-bias voltage corresponding to different RF driving voltages under 100 mTorr conditions obtained by iPM2D, experimental measurement \cite{hargis_gaseous_1994, overzet_microwave_1995}, and Rauf's model \cite{rauf_uniformity_2020} are presented in Fig.\ref{fig: 2}. The average electron number density and DC self-bias voltage calculated by iPM2D are in good agreement with the results of the experimental measurement and Rauf's model, thus verifying the implicit electrostatic PIC/MCC model in the 2D axisymmetric coordinate system developed in this paper. It should be explained that the low average electron number density calculated by iPM2D is due to the large time step used in the implicit push scheme mentioned in our previous report \cite{wang_implicit_2010}.

\begin{figure*}[ht]
    \centering
    \includegraphics[width=\linewidth]{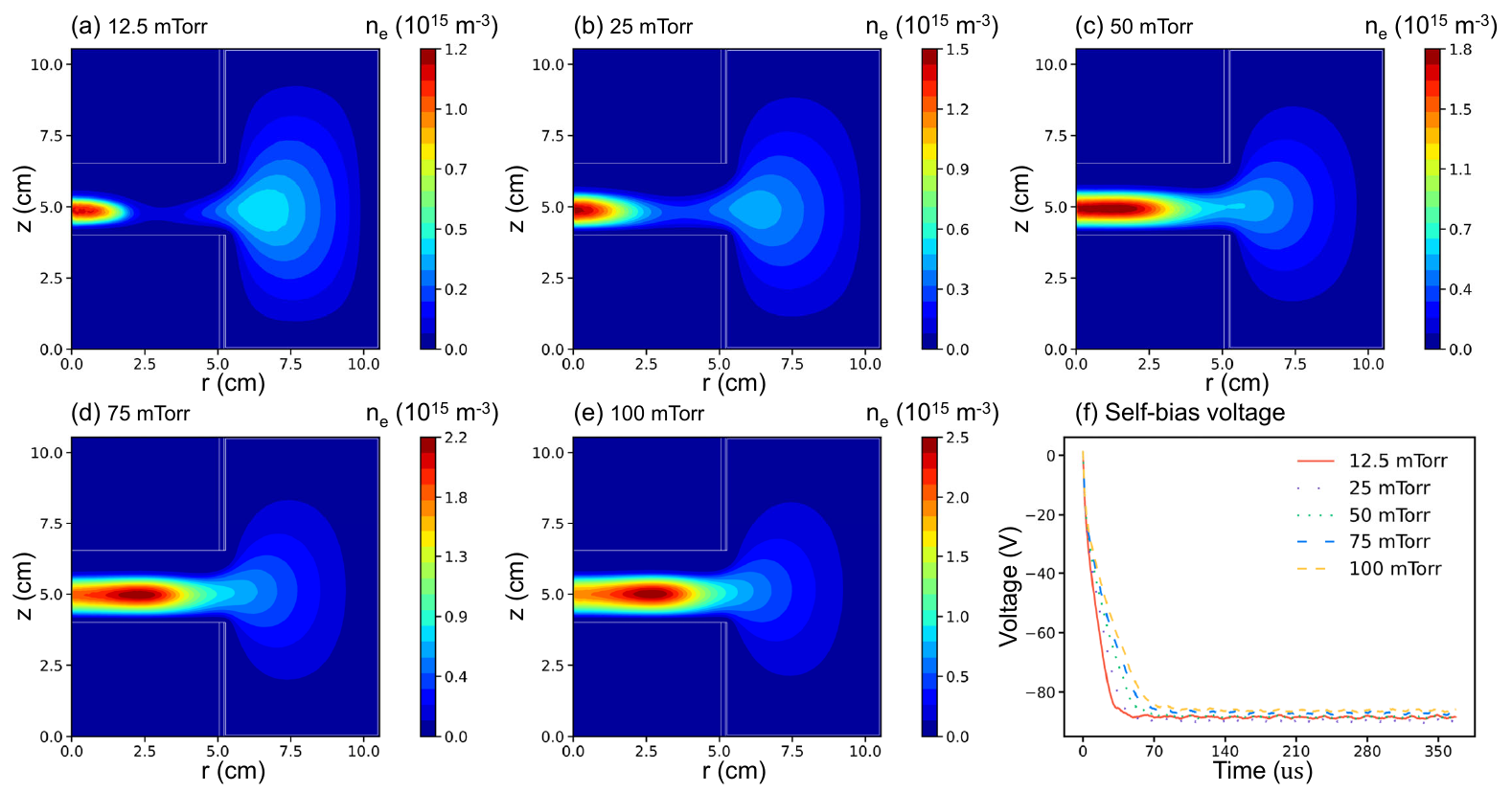}
    \caption{The spatial distribution of the electron number density at 12.5 mTorr (a), 25 mTorr (b), 50 mTorr (c), 75 mTorr (d) and 100 mTorr (e), and the self-bias voltage (f) changes with time at different pressures.}
    \label{fig: 3}
\end{figure*}

The present study provides spatially resolved results for the electron number density (averaged over 10 RF periods) for various pressures, as illustrated in Fig.\ref{fig: 3}. It is observed that an increase in pressure from 12.5 to 100 mTorr causes the plasma bulk to thicken, and the peak of the electron number density moves away from the central axis while increasing from $1.2 \times 10^{15} ~\rm m^{-3}$ to $2.5 \times 10^{15} ~\rm m^{-3}$. These findings agree with those Rauf reported \cite{rauf_uniformity_2020} and demonstrate the efficacy of the proposed model in simulating capacitively coupled discharges at low pressure while providing valuable insight into understanding plasma kinetics. Fig.\ref{fig: 3}-f shows the self-bias voltage curves at different pressures. The bias voltage on the top electrode gradually increases with time to balance the difference in plasma current density of the different electrodes caused by the asymmetry of the reaction geometry. By affecting the plasma current density, the pressure will also influence the establishment process and amplitude of self-bias.

\begin{figure*}[ht]
    \centering
    \includegraphics[width=\linewidth]{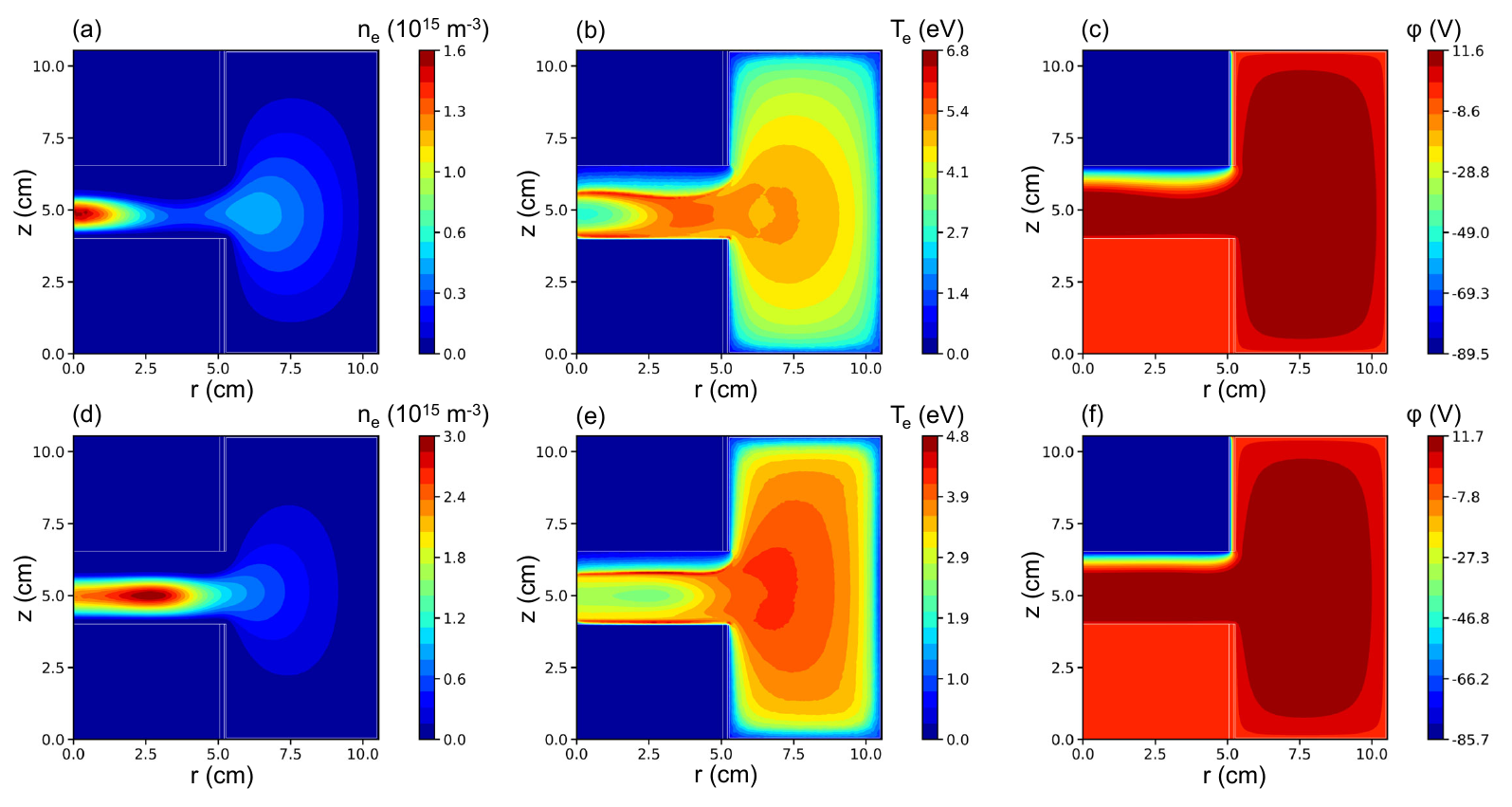}
    \caption{The spatial distribution of electron number density $n_e$, electron temperature $T_e$ and space potential $\phi$ at 25 mTorr (a, b, c) and 100 mTorr (d, e, f).}
    \label{fig: 4}
\end{figure*}

Circuit II is utilized to investigate the electrical characteristics of the GEC reference cell in the presence of external impedance matching. After 5000 RF cycles, the discharge reaches a steady state, and particle splitting is used during the last 2000 cycles. Under vacuum conditions, the RF driving voltage amplitude is set to 77 V, resulting in a voltage amplitude of 100 V on the top electrode of the GEC reactor. The simulation was carried out at various pressures ranging from 12.5 to 100 mTorr. Fig.\ref{fig: 4} shows the spatial distribution of the electron number density, the electron temperature, and the potential at 25 mTorr and 100 mTorr. At 25 mTorr, the mean free path of the electrons is larger, and the energetic electrons move freely throughout the chamber. The particle generation caused by ionization collision at each position in the chamber is roughly uniform, and the loss of particles to electrodes in the radial direction is the least at the central axis, resulting in the formation of a density peak at the central axis. At 100 mTorr, the increase in pressure leads to an increase of the electron collision frequency. The electron mean free path gets smaller and the diffusion of particles in the radial direction is limited, reducing the loss of particles. On the other hand, the electric field at the corner of the top and bottom electrodes in the GEC chamber is the strongest, where the energy of electrons obtained by collisionless heating is higher, and more particles are generated by ionization collisions locally. Therefore, a density peak is formed at r = 3 cm. A detailed discussion of plasma kinetics at low pressure is available in the literature \cite{rauf_uniformity_2020, kim_two-dimensional_2022}.

\begin{figure*}[ht]
    \centering
    \includegraphics[width=\linewidth]{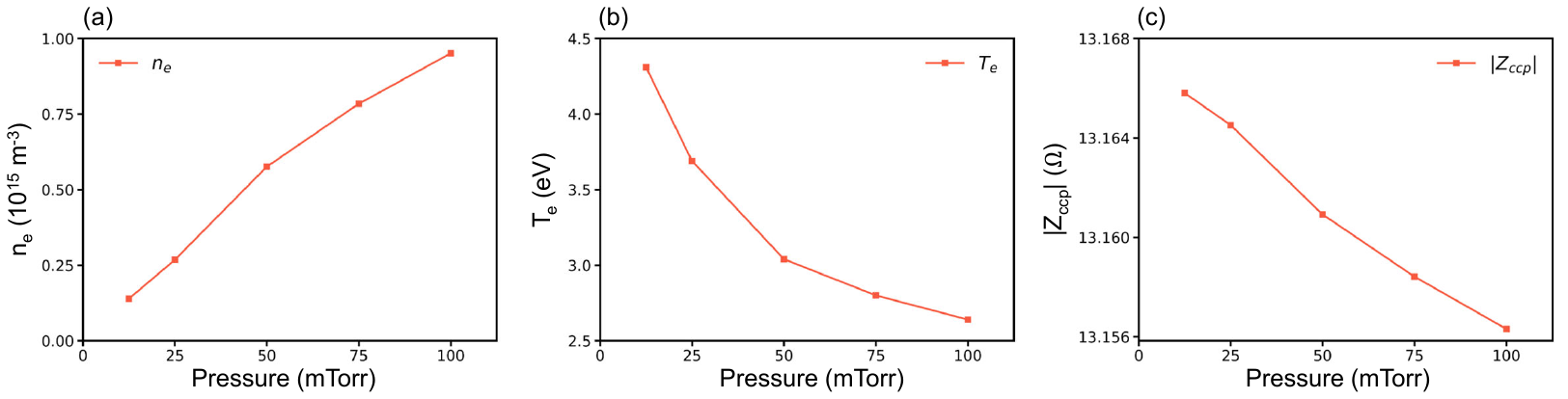}
    \caption{The average electron number density (a), the average electron temperature (b) and the module of equivalent impedance (c) of a CCP operated in the GEC reference cell at different pressures.}
    \label{fig: 5}
\end{figure*}

Tab.\ref{tab: power} presents results for the equivalent CCP impedance $Z_{ccp}$, the equivalent load impedance $Z_{load}$ (including IMN, stray branch and plasma load), the reflection coefficient $|\Gamma|$, the power consumption of each component, the reflected power $P_{ref}$, and the equivalent capacitance $C_{eq}$ of the GEC reactor at different pressures after discharge reaches steady state. The equivalent impedance is calculated using the fast Fourier transform (FFT) of the voltage $U_b$ and current $I_b$ components at the fundamental frequency as follows:
\begin{equation}
Z_{eq} = \frac{U_{b}}{I_{b}} \left[ cos \varphi + j sin \varphi \right]
\label{eq:equivalentcapacitance}
\end{equation}
where $\varphi$ is the phase difference between voltage and current at the fundamental frequency. The reflection coefficient $|\Gamma|$ is determined by the equivalent load impedance and the internal resistance $Z_s = R_s$ as follows:
\begin{equation}
\Gamma = \frac{Z_{load} - Z_s}{Z_{load} + Z_s}.
\label{eq:reflectioncoefficient}
\end{equation}
The active power consumed by each impedance is obtained from the periodic average power curve. T is the RF period.
\begin{equation}
P_{active} = \frac{\int_0^T U(t) I(t) dt}{T}
\label{eq: active power}
\end{equation}
The forward transmission power and the reflection coefficient determine the reflected power. Forward power is the power from the output of the power supply to the input of the impedance matching network.
\begin{equation}
P_{ref} = P_{forward} \times | \Gamma |^2
\label{eq: ref power}
\end{equation}
The calculation results show that the matching parameters calculated by the equivalent capacitance under vacuum conditions are very suitable and a small reflection coefficient can be achieved without multiple iterations. This results from the large capacitance of the GEC chamber. Under different pressures, even if the plasma state is different, the CCP equivalent impedance changes little. In this state, the total active power is approximately 30 W, of which almost half is consumed by the internal resistance $R_s$ of the power supply. This finding implies that the output power reaches the maximum under constant power supply voltage. The impedance matching network and the stray branch consume some of the output power, and almost 60 \% is absorbed by the plasma. The absorption ratio is consistent with the experimental results \cite{savas_dummy_1986, van_den_hoek_power_1987}. Fig.\ref{fig: 5} shows the average electron number density, the average electron temperature, and $Z_{ccp}$ at different pressures. It can be seen that the increase in pressure intensifies the collision processes, leading to an increase in the average electron number density and a decrease in the average electron temperature. The change in equivalent impedance is qualitatively consistent with the experimental results reported in \cite{kawata_power_1998}. The difference is that the capacitive characteristic of the GEC chamber is very strong and the impedance changes little.

\begin{table}[ht]
\centering
\caption{Electrical characteristic parameters at different pressures, 
including the CCP equivalent impedance $Z_{ccp}$, load equivalent impedances $Z_{load}$, reflection coefficient $|\Gamma|$, total power of the power supply $P_{source}$, Rs loss power $P_{Rs}$, IMN  loss power $P_{IMN}$, stray branch  loss power $P_{stray}$, CCP absorption power $P_{ccp}$, reflected power $P_{ref}$, and equivalent capacitance $C_{eq}$ of GEC.}
\label{tab: power}
\begin{tabular}{cccccc}
\hline
            & 12.5 mTorr            & 25 mTorr              & 50 mTorr              & 75 mTorr              & 100 mTorr             \\
\hline
$Z_{ccp} ~(\Omega)$   & 0.31-j13.16       & 0.31-j13.16       & 0.31-j13.16       & 0.31-j13.15       & 0.31-j13.15       \\
$Z_{load} ~(\Omega)$  & 48.96+j2.78       & 48.96+j2.65       & 49.03+j2.30       & 49.11+j2.08       & 49.17+j1.898       \\
$|\Gamma|$          & 0.030                & 0.029                & 0.025                & 0.023                & 0.021                \\
$P_{source} ~(W)$   & 29.9317               & 29.9349               & 29.9179               & 29.8960               & 29.8822               \\
$P_{Rs} ~(W)$       & 15.1229               & 15.1249               & 15.1051               & 15.0816               & 15.0664               \\
$P_{IMN} ~(W)$      & 5.8124                & 5.8102                & 5.8106                & 5.8118                & 5.8176                \\
$P_{stray} ~(W)$    & 0.0864                & 0.0864                & 0.0865                & 0.0861                & 0.0862                \\
$P_{ccp} ~(W)$      & 8.9100                & 8.9133                & 8.9156                & 8.9165                & 8.9119                \\
$P_{ref} ~(W)$      & 0.0133                & 0.0122                & 0.0094                & 0.0077                & 0.0064                \\
$C_{eq} ~(pF)$      & 892.17 & 892.28 & 892.54 & 892.72 & 892.86 \\
\hline
\end{tabular}
\end{table}

\begin{figure*}[ht]
    \centering
    \includegraphics[width=\linewidth]{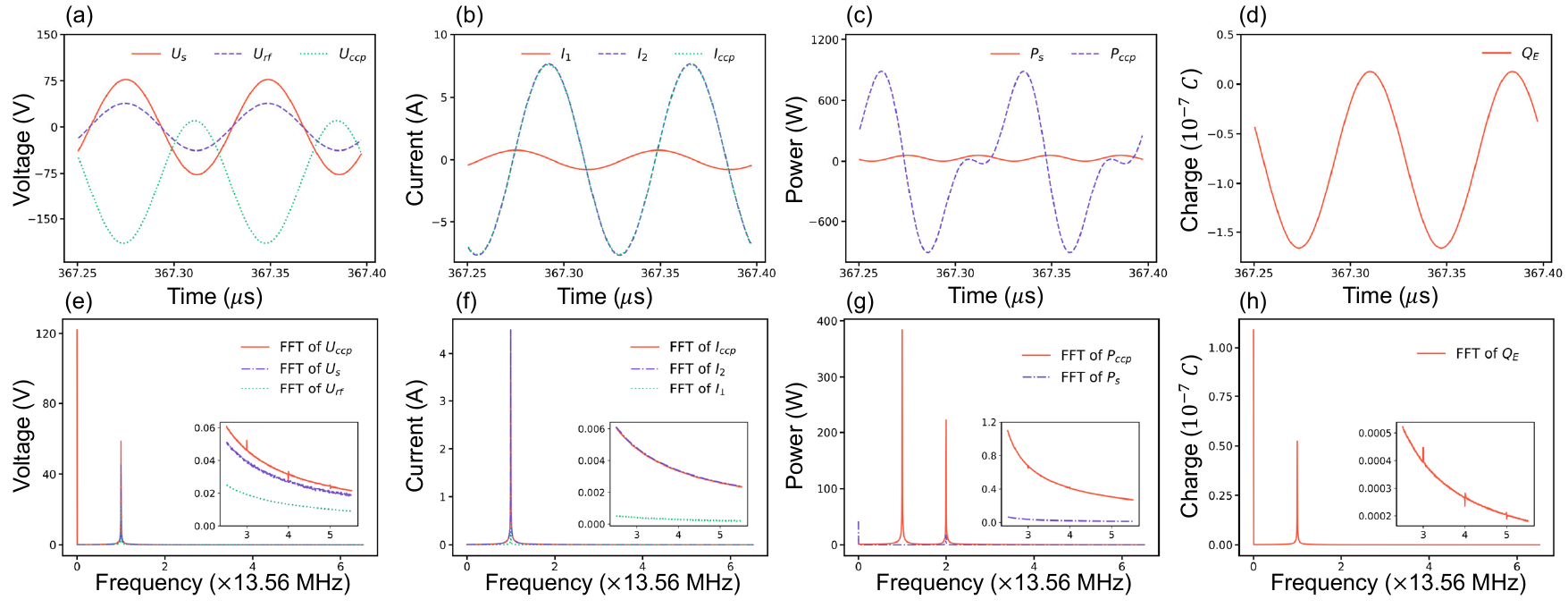}
    \caption{The waveforms of voltage (a), current (b), power (c) and charge (d) of the top electrode after the discharge reaches a steady state, as well as the fast Fourier transform results of voltage (e), current (f), power (g) and charge (h).}
    \label{fig: 6}
\end{figure*}

Fig.\ref{fig: 6} illustrates the waveforms of the voltage, current and power in the external circuit and the charge on the top electrode after the discharge reaches a steady state at 100 mTorr. Fast Fourier transform results are also presented. In this study, we denote $U_{rf} = \frac{Q_1 - Q_2}{C_{m1}}$ as the input voltage of the IMN. Under 77 V RF driving voltage, the peak-to-peak voltage value on the plasma load is 200 V, and the current amplitude is 7.6 A. In this matching state, the voltage and current in the load are higher than the RF input. In Fig. \ref{fig: 6}-b, the current $I_2$ almost coincides with $I_{ccp}$, indicating that the current in the stray branch is negligible. The FFT results of the voltage and current waveforms in Fig.\ref{fig: 6}-e and Fig.\ref{fig: 6}-f show that there are high-order harmonics, but the proportion is lower than 1 \%. The power waveform of the CCP shows a significant distortion. Fig.\ref{fig: 6}-g illustrates that the CCP power curve has a large component at 13.56 MHz, which distorts the power curve. Furthermore, the CCP power curve in Fig.\ref{fig: 6}-c shows that the discharge impedance of the GEC chamber exhibits a large capacitance characteristic. Although the amplitude of the power curve is high, its active power is very low. The amplitude of the power curve of the power supply is very small. However, it only has components at zero frequency and 27.12 MHz, indicating that the power supply power under the impedance matching condition is mostly active and the reflected reactive power is low. In Fig.\ref{fig: 6}-a and Fig.\ref{fig: 6}-d, the voltage and charge on the top electrode of the GEC reactor exhibit significant bias due to the strong asymmetry of the chamber.

\section{Conclusions} \label{sec4}
In summary, we presented a two-dimensional implicit electrostatic particle-in-cell/Monte Carlo collision (PIC/MCC) model named iPM2D, which enables the simultaneous solution of arbitrary external circuits and nonlinear plasmas in an axisymmetric coordinate system. The computed results of our model are validated against the results of experimental measurements \cite{hargis_gaseous_1994, overzet_microwave_1995} and Rauf's model \cite{rauf_uniformity_2020} and demonstrate excellent agreement. We utilized this model to investigate the characteristics of capacitively coupled discharges generated in the Gaseous Electronics Conference (GEC) reference cell with impedance matching. The simulation provides spatial distributions of plasma parameters, including density, temperature, and potential, accurately capturing plasma kinetics under low-pressure discharge conditions. Moreover, coupled with the external circuit, our model yields waveforms of circuit parameters, such as voltage, current, power, charge, and fast Fourier transform results, to study the electrical characteristics under impedance matching, and contributes significantly to understanding the nonlinear interaction between the external circuit and the plasma.


\section*{Acknowledgments}
This work was supported by the National Natural Science Foundation of China (12275095, 11975174, 12011530142, and 51821005).

\section*{Data availability statement}
The data that support the findings of this study are available from the corresponding author upon reasonable request.

\section*{ORCIDs}
Zili Chen https://orcid.org/0000-0002-2104-5369 \\
Shimin Yu https://orcid.org/0000-0002-3832-5252\\
Yu Wang https://orcid.org/0000-0003-3519-3677\\
Zhipeng Chen https://orcid.org/0000-0002-8330-0070\\
Wei Jiang https://orcid.org/0000-0002-9394-585X\\
J. Schulze https://orcid.org/0000-0001-7929-5734\\
Ya Zhang https://orcid.org/0000-0003-0473-467X\\

\nolinenumbers

\bibliography{library}

\bibliographystyle{abbrv}

\end{document}